\documentclass[twocolumn,showpacs,aps,prl,nofootinbib]{revtex4}

\newcommand{\be}{\begin{equation}}
\newcommand{\ee}{\end{equation}}
\newcommand{\la}{\lambda}

\newcommand{\bra}{\langle}
\newcommand{\ket}{\rangle}

\newcommand{\non}{\nonumber}
\newcommand{\bea}{\begin{eqnarray}}
\newcommand{\eea}{\end{eqnarray}}

\begin{document}

\title{Least paradoxical states of the Schr\"{o}dinger cat}
\author{Andrey Pereverzev}
\email{andrey.pereverzev@trinity.edu}
\affiliation{Department of Chemistry,
Trinity University, San Antonio, Texas 78212}

\date{\today}

\begin{abstract}
Modeling the Schr\"{o}dinger cat by a two state system 
 and assuming that the cat is coupled to the environment
we look for the least paradoxical states of the Schr\"{o}dinger cat
in the following way. We require the reduced density 
matrix of the cat  for
one of the two states in the superposition  to be the 
same as the one for the total state while 
distinct from the reduced density 
matrix of the cat for the other state in the superposition. 
We then look for the reduced density matrices for which
the cat is as alive as possible for the first state (and as dead as possible
for the second state). The resulting
states are those in which the probability for the cat to be
alive (or dead) is $1/2+\sqrt 2/4\approx 0.854$.
\end{abstract}
\maketitle

%%%%%%%%%%%%%%%%%%%%%%%%%%%%%%%%%%%%%%%%%%%%%%%%%%%%%%%%%%%%%%%%%%%%
In the thought experiment introduced by Schr\"{o}dinger \cite{Schrodinger} 
a superposition of microscopically distinct orthogonal states leads
to a superposition of macroscopically distinct orthogonal states of the type 
\be
|\chi\ket =\frac{1}{\sqrt 2}|\chi_1\ket +\frac{1}{\sqrt 2}|\chi_2\ket, 
\label{super}
\ee 
where $|\chi_1\ket$ describes the Universe in which the cat is alive,
while $|\chi_2\ket$ describes the Universe in which the cat is dead.
Rather then discussing possible resolutions of the paradox (see, e.g. 
\cite{Leggett}) we would like to 
consider for what states the paradoxical nature of superposition 
(\ref{super}) is least pronounced.
Let us denote the reduced density matrices of the cat for states
$|\chi\ket$, $|\chi_1\ket$, and $|\chi_2\ket$ by, respectively,
$\rho$, $\rho_1$, and $\rho_2$. The paradox would not exit if 
$\rho_1$ and $\rho_2$ corresponded, respectively,  to pure states for 
alive and dead  cat,  \emph{and} $\rho$ was identical to either 
$\rho_1$ or $\rho_2$. Clearly, this is not possible. However, 
since the cat can be entangled with the environment, we can relax
the requirement that  $\rho_1$ and  $\rho_2$ correspond to pure
states while keeping the requirement that $\rho$ is the same as 
either $\rho_1$ or  $\rho_2$. This would imply that the cat 
has some  probability of being dead for state $\rho_1$ and
some  probability of being alive for state $\rho_2$. Then we
can look for the density matrices $\rho_1$ and $\rho_2$ for
which the cat is, respectively, as alive and as dead as possible.
Our analysis show that the resulting states of the cat are those
for which its probability to be alive (or dead) is  
$1/2+\sqrt 2/4\approx 0.854$.

To formulate the problem more precisely
we model the cat by a two state system with states $|1\ket$ and
$|2\ket$ which correspond, respectively, to alive and dead states of the cat.
 We impose no restrictions on the form of the environment.
  Since the cat is modeled  by the
two state system  one can visualize the problem by using 
 the Bloch vector formalism \cite{Feynman}. 
Since we  require that $\rho$ is equal to, say, $\rho_1$,
 the Bloch 
vectors associated with $\rho$ and
$\rho_1$ have to be  the same.
 We also would like the reduced density matrices
 $\rho_1$ and  $\rho_2$ to correspond to  distinct states
of the cat. However, once $\rho_1$ is given, there is no unique way
to choose a distinct
$\rho_2$. One possible choice is based on the fact that in the Bloch 
picture two orthogonal pure states
 correspond to two vectors of unit length pointing in the opposite 
directions.   Generalizing this to the density matrices we will
require the vector for $\rho_2$ to be of the same length as for 
$\rho_1$  but pointing in the opposite direction.
 Let the Bloch vector for state $|1\ket$ 
to lie in the positive $z$ direction. 
Then our problem is to find vectors for  $\rho$,  $\rho_1$, and  
$\rho_2$ which, in addition to the requirements imposed above,
 have the largest absolute value of the $z$ component. 
It is clear from this picture that one necessary condition is that all 
three vectors are parallel to the $z$ axis.
To find the maximum possible length of the vectors and the corresponding 
probabilities it is convenient to write states $|\chi_1\ket$ and 
$|\chi_2\ket$ which appear in (\ref{super}) as the Schmidt decompositions, 
i. e., 
 \bea
|\chi_1\ket&=&\la|1\ket|\psi_1\ket+\sqrt{1-\la^2}|2\ket|\psi_2\ket, \non \\
|\chi_2\ket&=&\sqrt{1-\la^2}|1\ket|\phi_1\ket+\la|2\ket|\phi_2\ket. 
\label{Schmidt}
\eea
Here $\la^2$ is the probability that the cat is alive in state $|\chi_1\ket$
and that it is dead in state $|\chi_2\ket$. Since Eqs. (\ref{Schmidt}) are
the Schmidt decompositions, functions $|\psi_1\ket$ and  $|\psi_2\ket$  
are orthonormal as are functions  $|\phi_1\ket$ and  $|\phi_2\ket$.
Our analysis now reduces to the search for states $|\psi_1\ket$,
$|\psi_2\ket$, $|\phi_1\ket$, and $|\phi_2\ket$ which lead to 
the largest value for  $\la^2$ and  ensure that all the requirements
imposed above are satisfied.

Since we require states $|\chi_1\ket$ and $|\chi_2\ket$ to be orthogonal 
to each other we must have 
\be
\bra\phi_1|\psi_1\ket+\bra\phi_2|\psi_2\ket=
\bra\phi_1|\psi_1\ket^*+\bra\phi_2|\psi_2\ket^*=0. \label{cond1}
\ee
Substituting Eqs. (\ref{Schmidt}) into Eq. (\ref{super}) we obtain 
for $|\chi\ket$
\bea
|\chi\ket&=&|1\ket\left(\frac{\la}{\sqrt{2}}|\psi_1\ket
+\sqrt{\frac{1-\la^2}{2}}|\phi_1\ket\right)\non \\
& &+\,|2\ket\left(\sqrt{\frac{1-\la^2}{2}}|\psi_2\ket
+\frac{\la}{\sqrt{2}}|\phi_2\ket\right). \label{chicat}
\eea
As we want the reduced density matrix obtained from state $|\chi\ket$
to be the same as the one obtained from state $|\chi_1\ket$, state 
$|\chi\ket$  should 
have the Schmidt decomposition 
\be
|\chi\ket=\la|1\ket|\xi_1\ket+\sqrt{1-\la^2}|2\ket|\xi_2\ket. \label{chisch}
\ee
Comparing Eq. (\ref{chicat}) and Eq. (\ref{chisch}) we obtain for 
$|\xi_1\ket$ and $|\xi_2\ket$
\bea
|\xi_1\ket&=&\sqrt{\frac{1}{2}}|\psi_1\ket
+\sqrt{\frac{1-\la^2}{2\la^2}}|\phi_1\ket, \non \\
|\xi_2\ket&=&\sqrt{\frac{1}{2}}|\psi_2\ket
+\frac{\la}{\sqrt{2(1-\la^2)}}|\phi_2\ket.
\eea
For Eq.  (\ref{chisch}) to  be  the required Schmidt 
decomposition functions $|\xi_1\ket$ and $|\xi_2\ket$ have to be orthonormal.
It is sufficient to require that  $|\xi_1\ket$ and  $|\xi_2\ket$ are 
orthogonal and that one of them is normalized since conditions (\ref{cond1}) 
will ensure that the other state is also normalized. Requiring that 
state $|\xi_1\ket$
is normalized leads to the following equation for $\la$
\be
2\la^2-{A\la \sqrt{1-\la^2}}
-1=0, \label{lameq}
\ee
where $A=\bra\psi_1|\phi_1\ket+\bra\phi_1|\psi_1\ket$. 
The requirement of orthogonality of $|\xi_1\ket$ and $|\xi_2\ket$ leads to
\bea
& &(1-\la^2)\bra\phi_1|\psi_2\ket+\la^2\bra\psi_1|\phi_2\ket
=(1-\la^2)\bra\phi_1|\psi_2\ket^*\non \\
& &+\la^2\bra\psi_1|\phi_2\ket^*=0. \label{cond3}
\eea
Solving Eq. (\ref{lameq}) for $\la$ and keeping only the positive 
solution (since the Schmidt
decompositions allow only for positive expansion coefficients) 
we obtain
\be
\la=\sqrt{\frac{1}{2}-\frac{A}{2\sqrt{4+A^2}}}.
\ee
States $|\psi_1\ket$ and  $|\phi_1\ket$ are normalized, as a result
$A$ can take values between $-2$ and $2$. The maximum value of $\la$, 
$\la_m$ is
reached for $A=-2$. The corresponding probability $\la_{m}^2$ is 
given by
\be
\la_{m}^2=\frac{1}{2}+\frac{\sqrt 2}{4} \approx 0.854. \label{lamax}
\ee
For $A$ to be equal to $-2$ we must have $|\phi_1\ket=-|\psi_1\ket$.
Condition (\ref{cond1}) then requires that $|\phi_2\ket=|\psi_2\ket$. The 
last two equalities ensure that condition (\ref{cond3}) is also satisfied.

Thus, states  $|\chi\ket$, $|\chi_1\ket$, and $|\chi_2\ket$ with the 
sought after
properties will have the form
\bea
|\chi\ket&=&
\la_{m}|1\ket|\psi_1\ket-\sqrt{1-\la_{m}^2}|2\ket|\psi_2\ket, \non \\
|\chi_1\ket&=&
\la_{m}|1\ket|\psi_1\ket+\sqrt{1-\la_{m}^2}|2\ket|\psi_2\ket, \non \\
|\chi_2\ket&=&
\sqrt{1-\la_{m}^2}|1\ket|\psi_1\ket-\la_{m}|2\ket|\psi_2\ket, 
\label{states}
\eea
where $|\psi_1\ket$ and $|\psi_2\ket$ are two 
arbitrary orthonormal states of the environment and $\la_{m}$ 
is obtained from (\ref{lamax}). To sum up, the reduced
density matrices of the cat for states $|\chi\ket$ and  
$|\chi_1\ket$ are identical and diagonal in the basis of
states $|1\ket$ and
$|2\ket$, the probability for the cat to be alive is
about  0.854. The reduced density matrix of the cat for 
state $|\chi_2\ket$ is also diagonal in the same basis, 
the probability for the cat to be dead is
about  0.854. 

It is easy to see that the limit $\la_{m}$ appears already on the level
of an isolated two state system. Consider the following orthonormal 
states for two state system
\bea 
|I\ket=\la_{m}|1\ket+\sqrt{1-\la_{m}^2}|2\ket, \non \\
|II\ket=\sqrt{1-\la_{m}^2}|1\ket-\la_{m}|2\ket,
\eea
and their superposition
\be
|III\ket=\frac{1}{\sqrt 2}|I\ket+\frac{1}{\sqrt 2}|II\ket
=\la_{m}|1\ket|-\sqrt{1-\la_{m}^2}|2\ket.
\ee
Note that states $|I\ket$,  $|II\ket$, and  $|III\ket$ have
the same probabilities for the cat to be alive as, respectively,
states $|\chi_1\ket$, $|\chi_2\ket$, and $|\chi\ket$.
If we use  the Bloch vector formalism and chose the vector for 
state $|1\ket$ to lie  along the 
positive $z$ axis, then state $|I\ket$
corresponds to vector ${\bf{P}}_I$ at $45$ degrees to the $z$ axis.
The Bloch vector  ${\bf{P}}_{II}$ for state $|II\ket$ is of the same length as 
${\bf{P}}_I$ but pointing in the opposite direction, 
while vector ${\bf{P}}_{III}$ for state $|III\ket$ 
is at $45$ degrees to the $z$ axis and perpendicular to both ${\bf{P}}_I$ 
and ${\bf{P}}_{II}$.
It is clear from this picture that vectors ${\bf{P}}_I$ and ${\bf{P}}_{II}$ 
will have $z$ components equal to $\sqrt 2/2$
with the corresponding probability for the cat to be alive 
given by $\la^2_{m}$ while the $z$ component for vector ${\bf{P}}_{III}$ 
will be $-\sqrt 2/2$. 
By using this geometrical picture one can convince oneself that this 
arrangement of vectors insures the largest absolute value of 
the $z$ component for the three vectors. All three vectors, however, will 
have non zero and
different $x$ and $y$ components. 
Thus, coupling the cat to the environment in 
states (\ref{states}) cannot increase the absolute value of the $z$ 
components and the
corresponding probabilities. However, it allows to 
 remove the $x$ and $y$ components of the
Bloch vectors for all three states making the reduced density matrices 
obtained from states $|\chi\ket$ and $|\chi_1\ket$ identical.
It can be argued that in the resulting states (\ref{states}) the fate 
of the cat is less dramatic than in the original 
formulation of the paradox.

\begin{acknowledgments}
The author would like to thank Dr. Gonzalo Ord\'{o}\~{n}ez for
useful discussions. Part of this research was supported by grants
from the Robert A. Welch Foundation (Grant No. W-1442) and the Petroleum
Research Fund, administered by the American Chemical Society.
\end{acknowledgments}

\end{document}